# QUANTUM COMMUNICATIONS BASED ON QUANTUM HASHING

Alexander Vasiliev

*Kazan Federal University*

**Abstract:** In this paper we consider an application of the recently proposed quantum hashing technique for computing Boolean functions in the quantum communication model. The combination of binary functions on non-binary quantum hash function is done via polynomial presentation, which we have called a characteristic of a Boolean function.

Based on the characteristic polynomial presentation of Boolean functions and quantum hashing technique we present a method for computing Boolean functions in the quantum one-way communication model, where one of the parties performs his computations and sends a message to the other party, who must output the result after his part of computations. Some of the results are also true in a more restricted Simultaneous Message Passing model with no shared resources, in which communicating parties can interact only via the referee.

We give several examples of Boolean functions whose polynomial presentations have specific properties allowing for construction of quantum communication protocols that are provably exponentially better than classical ones in the simultaneous message passing setting.

## 1. INTRODUCTION

While a large-scale fully functional quantum computer remains a theoretical model, quantum communications are extensively implemented and may soon enter our daily life. That is why the study of different quantum communication models could add value to this technology. However, in absence of long-term quantum memory and quite small coherence time of quantum states we should consider restricted versions of quantum communication models in the first place. In particular, such models include those considered here: the one-way quantum communication model and the more restricted simultaneous message passing (SMP) model [1] with no shared resources.

From the complexity theoretic viewpoint such a strong restrictions on a computational model allow a variety of techniques for proving lower bounds on the complexity in this model. Sometimes it even allows to prove exponential separations for quantum and classical models, e.g. Buhrman et al. [2] proposed a fingerprinting-based protocol for computing Equality in the SMP model that uses $O(\log n)$ quantum bits of communication while in classical case it requires $\Omega(\sqrt{n})$. Besides, restricted communication models have proved their usefulness in proving lower bounds for related models (see, for example, [3], [4], [5]).

In this paper we focus on proving upper bounds for a class of Boolean functions, described by the properties of their polynomial presentations. Our approach relies on the polynomial presentation of Boolean functions, which has proven its usefulness in a number of papers [6], [7], [8]. However, here we use a slightly different type of polynomial presentation proposed in [9].

Another component of our approach is quantum hashing [10], which transforms a classical input into quantum superposition. In [10] we have shown that quantum hashing can have applications in quantum cryptography and in [11] we have demonstrated computational aspect of this technique. Here, hashing is used to reduce the amount of data transferred between communicating parties, just like it has been done by means of quantum fingerprinting in [2]. However, we have proposed that quantum fingerprinting is also a quantum hashing in terms of [10].

Altogether, the main construction provides effective one-way quantum communication protocols for the class of functions with specific polynomial presentation. This construction can be immediately used in the SMP model, however, when this construction is generalized for arbitrary Boolean function it is valid for one-way model only. We show that several known Boolean functions are in this class.

### 2. PRELIMINARIES

At the core of our approach lies the polynomial presentation of Boolean functions proposed in [9]. We recall some of the definitions here.

**Polynomial presentation of Boolean functions.**

**Definition 2.1.** *We call a polynomial $g(x_1, \ldots, x_n)$ over the ring $\mathbb{Z}_m$ a characteristic polynomial of a Boolean function $f(x_1, \ldots, x_n)$ and denote it $g_f$ when for all $\sigma \in \{0,1\}^n$ it holds $g_f(\sigma) = 0$ iff $f(\sigma) = 1$.*

It was also shown that such a polynomial always exists (but is not unique).

**Lemma 2.1.** *For any Boolean function $f$ of $n$ variables there exists a characteristic polynomial $g_f$ over $\mathbb{Z}_{2^n}$.*

*Proof.* One way to construct such characteristic polynomial $g_f$ is transforming a sum of products representation for $\neg f$.

Let $K_1 \vee \ldots \vee K_l$ be a sum of products for $\neg f$ and let $\widetilde{K}_i$ be a product of terms from $K_i$ (negations $\neg x_j$ are replaced by $1 - x_j$). Then $\widetilde{K}_1 + \cdots + \widetilde{K}_l$ is a characteristic polynomial over $\mathbb{Z}_{2^n}$ for $f$ since it equals 0 iff all of $\widetilde{K}_i$ (and thus $K_i$) equal 0. This happens only when the negation of $f$ equals 0.

Generally, there are many polynomials for the same function. For example, the function $EQ_n$, which tests the equality of two n-bit binary strings, has the following polynomial over $\mathbb{Z}_{2^n}$:

$$\sum_{i=1}^{n}(x_i(1-y_i) + (1-x_i)y_i) = \sum_{i=1}^{n}(x_i + y_i - 2x_iy_i).$$

On the other hand, the same function can be represented by the polynomial

$$\sum_{i=1}^{n} x_i 2^{i-1} - \sum_{i=1}^{n} y_i 2^{i-1}.$$

We have used this presentation to test a single property of the input encoded by a characteristic polynomial. Using the same ideas we can test the conjunction of several conditions encoded by a group of characteristic polynomials which we call a *characteristic* of a function.

**Definition 2.1.** *We call a set $\chi_f$ of polynomials over $\mathbb{Z}_m$ a characteristic of a Boolean function $f$ if for any $\sigma \in \{0,1\}^n$ it holds that all polynomials $g \in \chi_f$ equal 0 on $\sigma$ iff $\sigma \in f^{-1}(1)$.*

From Lemma 2.1 it follows that for each Boolean function there is always a characteristic consisting of a single characteristic polynomial.

We say that a characteristic is *linear* if all of its polynomials are linear. In [9] we have shown that Boolean functions with linear characteristics of logarithmic size can be efficiently computed in the quantum OBDD model.

**Quantum communication protocols**. Quantum communication protocol is a generalization of randomized communication protocols (see for example [4]). In a quantum protocol both players (conventionally called Alice and Bob) have a private set of qubits each initialized to $|0\rangle$. At the beginning both receive a Boolean string that encodes their part of the input. In each communication round one player applies a unitary transformation to the qubits in his possession (including those received from the other communicating party) and then sends some of the qubits to the other player. At the end of the protocol the state of some qubits belonging to one player is measured and he outputs the result (the value of the function being computed).

The one-way restriction implies a single round of communication, when Alice sends some information to Bob, who outputs the result.

We also consider a restricted variant of quantum communication model known as simultaneous message passing (SMP) model [1] with no shared resources, which involves three communicating parties: Alice, Bob and referee. Alice and Bob do not interact directly. Instead, they send their messages to the referee during a single round, and the referee must output the result.

In each case the complexity of the protocol is the amount of qubits sent between the parties.

**Quantum hashing.** We recall a definition of quantum hashing function from [10].

Let $q = 2^n$ and $B = \{b_1, b_2, \ldots, b_d\} \subset \mathbb{Z}_q$. We define a quantum hash function $\psi_{q,B}: \{0,1\}^n \to (\mathcal{H}^2)^{\otimes(\log d + 1)}$ as follows. For an input $x \in \{0,1\}^n$ we let

$$|\psi_{q,B}(x)\rangle = \frac{1}{\sqrt{d}} \sum_{i=1}^{d} |i\rangle \left( \cos\frac{2\pi b_i x}{q} |0\rangle + \sin\frac{2\pi b_i x}{q} |1\rangle \right).$$

It follows from this definition that the quantum hash $|\psi_{q,B}(x)\rangle$ of an $n$-bit string $x$ consists of $\log d + 1$ qubits. The set $B = \{b_1, b_2, \ldots, b_d\}$ of hashing parameters not only determines the size of the hash but also gives the function $\psi_{q,B}$ an ability to withstand collisions, i.e. to distinguish different hashes with bounded error probability. We have called this property $\delta$-*resistance*.

Formally, for $\delta \in (0,1)$ we call a function $\psi: \mathbb{X} \to (\mathcal{H}^2)^{\otimes s}$ $\delta$-resistant if for any pair $w, w'$ of different inputs

$$|\langle\psi(w)|\psi(w')\rangle| \leq \delta.$$

We have shown that $d$ can be of order $O(n)$ without losing the quality of hashing [10].

Thus, for a quantum hash-function it is important to have an ability to reliably compare quantum hashes of different words and those quantum states need to be distinguishable with high probability, that is, they have to pass non-equality tests.

**REVERSE-test.** Whenever we need to check if a quantum state $|\psi(w)\rangle$ is a hash of a classical message $v$, one can use the procedure that we have called a *REVERSE-test*.

Essentially the test applies the procedure that inverts the creation of a quantum hash, i.e. it "uncomputes" the hash to the initial state (usually the all-zero state).

Formally, let the procedure of quantum hashing of message w consist of unitary transformation $U(w)$, applied to initial state $|0\rangle$, i.e. $|\psi(w)\rangle = U(w)|0\rangle$. Then the REVERSE-test, given $v$ and $|\psi(w)\rangle$, applies $U^{-1}(v)$ to the state $|\psi(w)\rangle$ and measures the resulting state. It outputs $v = w$ iff the measurement outcome is $|0\rangle$. So, if $v = w$, then $U^{-1}(v)|\psi(w)\rangle$ would always give $|0\rangle$, and REVERSE-test would give the correct answer. Otherwise, the probability of error would be bounded by $\delta$ by $\delta$-resistance of the hash function [10].

**SWAP-test.** A more general test, that checks the equality of two arbitrary states is a well-known SWAP-test [2], given by the circuit in Fig. 1.

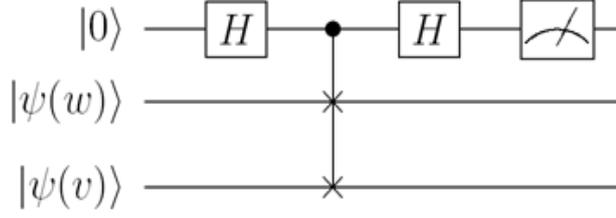

**Fig. (1).** A quantum circuit for SWAP-test.

Applied to quantum hashes it outputs $|\psi(w)\rangle = |\psi(v)\rangle$, if the measurement result of the first qubit is $|0\rangle$.

**Property 2.1.** *The probability of obtaining $|0\rangle$ in the SWAP-test is equal to $(1+|\langle\psi(w)|\psi(v)\rangle|^2)/2$.*

The probability of error of the SWAP-test inherently depends on the value of the inner product of $|\psi(w)\rangle$ and $|\psi(v)\rangle$, i.e. on the $\delta$-resistance of the underlying quantum hash function.

## 3. QUANTUM COMMUNICATION PROTOCOLS BASED ON QUANTUM HASHING

The quantum hashing defined above can be used for constructing effective protocols in the quantum communication model defined by Yao in [12].

Here we consider a one-sided restriction of this model, where Alice makes her computations, sends some information to Bob, who computes his part of the protocol and outputs the result. The complexity of such a protocol is the number of qubits sent to Bob.

Let $f(x_1, \ldots, x_{n_1}, y_1, \ldots, y_{n_2})$ be a Boolean function of $n = n_1 + n_2$ variables, i.e.

$$f:\{0,1\}^{n_1} \times \{0,1\}^{n_2} \to \{0,1\}.$$

Alice gets the sequence of values $\sigma = \sigma_1 \ldots \sigma_{n_1}$ of the first $n_1$ variables, and Bob gets $\gamma = \gamma_1 \ldots \gamma_{n_2}$ - the values of the last $n_2$ variables.

To compute $f$ we exploit its polynomial presentation, described in Section 2.

In the communication scenario the input is split between parties, and a polynomial for f should also be decomposed. For the quantum hashing technique proposed we decompose this polynomial into the sum of two polynomials, one for each of the communicating parties.

**Theorem 3.1.** *Let $f(x_1, \ldots, x_{n_1}, y_1, \ldots, y_{n_2})$ be a Boolean function of $n = n_1 + n_2 2$ variables. Let $g$ be a characteristic polynomial for $f$ over the ring $\mathbb{Z}_q$. If $g$ can be decomposed into*

$$g(x_1, \ldots, x_{n_1}, y_1, \ldots, y_{n_2}) = g_1(x_1, \ldots, x_{n_1}) + g_2(y_1, \ldots, y_{n_2}),$$

*then for arbitrary $\delta \in (0,1)$ $f$ can be computed by a one-way quantum communication protocol with $O(\log\log q + \log(1/\delta))$ qubits of communication.*

*Proof.* For the proof we describe the following quantum one-way communication protocol.

The communicating parties given an input $(\sigma, \gamma)$ want to know whether $f(\sigma, \gamma) = 1$ or not. This is the same as asking whether $g(\sigma, \gamma) = 0$, or, equivalently, whether $g_1(\sigma_1, \ldots, \sigma_{n_1}) = -g\_2(\gamma_1, \ldots, \gamma_{n_2})$. And this equality is exactly what the protocol would check using quantum hashing technique, i.e. it will compare quantum hashes of those values.

More formally, the following describes a one-way protocol of computing $f$ in the quantum communication setting using $\delta$-resistant quantum hashing for some $\delta \in (0,1)$.

1. Alice, depending on her input $\sigma = \sigma_1 \ldots \sigma_{n_1}$, creates a quantum hash for the value $g_1(\sigma)$

$$|\psi_{q,B}(g_1(\sigma))\rangle = \frac{1}{\sqrt{d}} \sum_{i=1}^{d} |i\rangle \left( \cos \frac{2\pi b_i g_1(\sigma)}{q} |0\rangle + \sin \frac{2\pi b_i g_1(\sigma)}{q} |1\rangle \right)$$

and sends it to Bob.

2. Given $|\psi_{q,B}(g_1(\sigma))\rangle$ and his input $\gamma = \gamma_1 \ldots \gamma_{n_2}$ Bob creates a quantum hash for the value $-g_2(\gamma)$

$$|\psi_{q,B}(-g_2(\gamma))\rangle = \frac{1}{\sqrt{d}} \sum_{i=1}^{d} |i\rangle \left( \cos \frac{2\pi b_i (-g_2(\gamma))}{q} |0\rangle + \sin \frac{2\pi b_i (-g_2(\gamma))}{q} |1\rangle \right)$$

3. Bob compares $|\psi_{q,B}(g_1(\sigma))\rangle$ and $|\psi_{q,B}(-g_2(\gamma))\rangle$ using the SWAP-test. So, Bob obtains the result $|\psi_{q,B}(g_1(\sigma))\rangle = |\psi_{q,B}(-g_2(\gamma))\rangle$, if the measurement of the first qubit gives $|0\rangle$, which happens with probability $\left(1 + |\langle\psi_{q,B}(g_1(\sigma))|\psi_{q,B}(-g_2(\gamma))\rangle|^2\right)/2$.

4. Bob outputs the result of computations. He says $f(\sigma, \gamma) = 1$ if $|\psi_{q,B}(g_1(\sigma))\rangle = |\psi_{q,B}(-g_2(\gamma))\rangle$ and $f(\sigma, \gamma) = 0$ otherwise. If the value of $f(\sigma, \gamma)$ was 1, then Bob outputs 1 with certainty. If $f(\sigma, \gamma)$ was 0, then by $\delta$-resistance property $|\langle\psi_{q,B}(g_1(\sigma))|\psi_{q,B}(-g_2(\gamma))\rangle| \leq \delta$, and the probability of erroneously outputting 1 is bounded by $1/2 + \delta^2/2$.

The communication complexity in this case is bounded by the size of the quantum hash passed from Alice to Bob, which is $\log d + 1 = O(\log \log q + \log 1/\delta)$ qubits.

**Remark 3.1.** By inspecting the proposed communication protocol one can note that if instead of directly communicating Alice and Bob had to send their hashes to the referee, we would have a protocol in a more restrictive setting of *simultaneous message passing model*, and the Theorem 3.1 can be restated and proved for this model as well.

**Remark 3.2.** The probability of error in the construction of Theorem 3.1 can be reduced to $\delta$ if Bob would perform REVERSE-test of the received quantum hash and his computed value $-g_2(\gamma)$. Unfortunately, this is not the case for the simultaneous message passing model.

### 3.1. SOME EXAMPLES

The theorem above assumes that characteristic polynomial can be decomposed into the sum of polynomials over independent sets of variables. The simplest case of such polynomials is linear polynomials and we have exposed in [9] several examples of natural Boolean functions that have linear characteristic polynomials. Among them there is an *Equality test*, which is frequently considered in the study of communication complexity. The corresponding Boolean function has the following linear characteristic polynomial over $\mathbb{Z}_{2^n}$

$$g_{EQ}(x_1, \ldots, x_n, y_1, \ldots, y_n) = \sum_{i=1}^{n} x_i 2^{i-1} - \sum_{i=1}^{n} y_i 2^{i-1}$$

and thus can be computed by the $O(\log n)$-qubit quantum communication protocol.

Here are some more Boolean functions with linear characteristic polynomials, which are thus effectively computable in the SMP model.

$MOD_m$  The function $MOD_m$ tests whether the number of 1's in the input is 0 modulo $m$. The linear polynomial over $\mathbb{Z}_m$ for this function is

$$\sum_{i=1}^{n} x_i .$$

$MOD'_m$  This function is the same as $MOD_m$, but the input is treated as a binary number. Thus, the linear polynomial is

$$\sum_{i=1}^{n} x_i 2^{i-1} .$$

$Palindrome_n(x_1, \ldots, x_n)$   This function tests the symmetry of the input, i.e. whether $x_1 x_2 \ldots x_{\lfloor n/2 \rfloor} = x_n x_{n-1} \ldots x_{\lceil n/2 \rceil+1}$ or not. The polynomial over $\mathbb{Z}_{2^{\lfloor n/2 \rfloor}}$ is

$$\sum_{i=1}^{\lfloor n/2 \rfloor} x_i 2^{i-1} - \sum_{i=\lceil \frac{n}{2} \rceil}^{n} x_i 2^{n-i} .$$

$PERM_n$  The *Permutation Matrix* test function ($PERM_n$) is defined on $n^2$ variables $x_{ij}$ ($1 \leq i, j \leq n$). It tests whether the input matrix contains exactly one 1 in each row and each column. Here is a polynomial over $\mathbb{Z}_{(n+1)^{2n}}$

$$\sum_{i=1}^{n}\sum_{j=1}^{n} x_{ij}\left((n+1)^{i-1} + (n+1)^{n+j-1}\right) - \sum_{i=1}^{2n} (n+1)^{i-1} .$$

### 3.2. AN EXTENSION FOR ARBITRARY BOOLEAN FUNCTION

Now, if for some Boolean function $f$ there is no characteristic polynomial, that can be decomposed as shown earlier, we use the following decomposition

$$g(x_1, \ldots, x_{n_1}, y_1, \ldots, y_{n_2}) = g_1(x_1, \ldots, x_{n_1}) + g_2(x_{i_1}, \ldots, x_{i_k}, y_1, \ldots, y_{n_2}) .$$

Such a decomposition always exists, since we can choose $k = n_1$, $g_1 \equiv 0$ and $g_2 \equiv g$.

Then the following result holds, which generalizes Theorem 3.1.

**Theorem 3.2.** *For arbitrary $\delta \in (0,1)$ $f$ can be computed by a one-way quantum communication protocol with $O(k + \log\log q + \log(1/\delta))$ qubits of communication.*

*Proof.* The protocol is almost the same as the one from Theorem 3.1, but Alice sends a hash plus $k$ qubits containing the values of $x_{i_1}, \ldots, x_{i_k}$, and Bob use them to construct his own hash. The protocol now requires $O(k + \log\log q + \log(1/\delta))$ qubits of communication.

**Corollary 3.1.** *If $q = 2^{n^{O(1)}}$ (which is the most usual case) and $k = O(\log n)$ the described protocol would require $O(\log n)$ qubits of communication, which is exponentially better than just sending all of the input from Alice to Bob.*

**Remark 3.3.** However, for arbitrary Boolean function the bound is still no better than trivial $O(n)$, since in general $k = O(n)$.

## 4. GENERAL APPROACH

In a more general approach, for some Boolean function $f$ depending on $n = n_1 + n_2$ variables we consider a characteristic $\chi_f^q$ over $\mathbb{Z}_q$.

Let us pick two sets $G = \{g_1, \ldots, g_l\}$ and $R = \{r_1, \ldots, r_l\}$ of polynomials over the ring $\mathbb{Z}_q$, such that the set $\chi_f^q = \{g_1 + r_1, \ldots, g_l + r_l\}$ is a characteristic of $f$ over $\mathbb{Z}_q$. Here we assume that polynomials from $G$ depend only on $x_1, \ldots, x_{n_1}$, and those from $R$ – depend not only on $y_1, \ldots, y_{n_2}$, but also on $x_{i_1}, \ldots, x_{i_k}$. Then the following theorem holds.

Theorem 4.1. For arbitrary $\delta \in (0,1)$ $f$ can be computed by a one-way quantum communication protocol with $O\big(k + |\chi_f^q| + \log\log q + \log(1/\delta)\big)$ qubits of communication.

*Proof.* The proof below constructs a protocol that is some generalization of the one from Theorem 3.1.

1. The protocol starts when Alice receives an input $\sigma = \sigma_1 \ldots \sigma_{n_1}$, combines the values $g_1(\sigma), \ldots, g_l(\sigma)$ into the following generalized quantum hash of $\log d + l$ qubits

$$|\psi_{q,B}(G(\sigma))\rangle = \frac{1}{\sqrt{d}} \sum_{j=1}^{d} |j\rangle \left(\cos\frac{2\pi b_j(g_1(\sigma))}{q}|0\rangle + \sin\frac{2\pi b_j(g_1(\sigma))}{q}|1\rangle\right) \ldots$$
$$\ldots \left(\cos\frac{2\pi b_j(g_l(\sigma))}{q}|0\rangle + \sin\frac{2\pi b_j(g_l(\sigma))}{q}|1\rangle\right),$$

and sends it to Bob along with the values $x_{i_1}, \ldots, x_{i_k}$.

2. Bob receives his part of the input $\gamma = \gamma_1 \ldots \gamma_{n_2}$, the quantum hash and values $x_{i_1}, \ldots, x_{i_k}$ from Alice. Then he computes his own hash for $-r_1(\gamma), \ldots, -r_l(\gamma)$

$$|\psi_{q,B}((-R(\gamma)))\rangle = \frac{1}{\sqrt{d}} \sum_{j=1}^{d} |j\rangle \left( \cos \frac{2\pi b_j(-r_1(\gamma))}{q} |0\rangle + \sin \frac{2\pi b_j(-r_1(\gamma))}{q} |1\rangle \right) \ldots$$
$$\ldots \left( \cos \frac{2\pi b_j(-r_l(\gamma))}{q} |0\rangle + \sin \frac{2\pi b_j(-r_l(\gamma))}{q} |1\rangle \right),$$

and performs the SWAP-test.

3. Bob outputs 1 iff all the hashes have passed the test, which happens with probability $\left(1 + |\langle \psi_{q,B}(G(\sigma))|\psi_{q,B}(-R(\gamma))\rangle|^2\right)/2$, which equals

$$\frac{1}{2} + \frac{1}{2d^2} \left( \sum_{j=1}^{d} \cos \frac{2\pi b_j(g_1(\sigma) + r_1(\gamma))}{q} \ldots \cos \frac{2\pi b_j(g_l(\sigma) + r_l(\gamma))}{q} \right)^2.$$

When $f(\sigma, \gamma) = 1$, this protocol would always lead to correct results. But if $f(\sigma, \gamma) = 0$, then for at least one $i \in \{1, \ldots, l\}$ $g_i(\sigma) \neq -r_i(\gamma)$, and the probability of outputting 1 is bounded by

$$\frac{1}{2} + \frac{1}{2d^2} \left( \sum_{j=1}^{d} \cos \frac{2\pi b_j(g_1(\sigma) + r_1(\gamma))}{q} \ldots \cos \frac{2\pi b_j(g_l(\sigma) + r_l(\gamma))}{q} \right)^2 \leq$$

$$\frac{1}{2} + \frac{1}{2d^2} \left( \sum_{j=1}^{d} \cos \frac{2\pi b_j(g_i(\sigma) + r_i(\gamma))}{q} \right)^2 \leq \frac{1}{2} + \frac{\delta^2}{2}.$$

The last inequality is based on the $\delta$-resistance property proof in [10].

Thus, the complexity of communication protocols based on quantum hashing and general characteristic polynomial presentation of Boolean functions is $O(k + |\chi_f^q| + \log \log q + \log(1/\delta))$.

**Corollary 4.1.** *Whenever $k = O(\log n)$, $|\chi_f^q| = O(\log n)$, and $q = 2^{n^{O(1)}}$, the complexity of such protocol would be $O(\log n)$.*

An example of such function is a Boolean version of Hidden Subgroup Problem, considered in [9], which has a characteristic over $\mathbb{Z}_{2^n}$, consisting of two polynomials.

## 5. SUMMARY

To summarize, we have proposed an approach for constructing quantum communication protocols for Boolean functions given by their characteristic polynomial presentation. Quantum hashing technique is used here to reduce the amount of information being sent between communicating parties.

## 6. CONCLUSIONS

The construction presented in this paper uses quantum hashing technique for distributed quantum computations. The communication complexity of Boolean functions being computed depends on the properties of their polynomial presentation and collision resistance of the underlying quantum hash function. Generally, any Boolean function can be computed using this approach, but the complexity bound would be trivial. However, for certain classes of Boolean functions with "good" polynomial presentation the resulting quantum communication protocols are exponentially better than their classical counterparts.

**ACKNOWLEDGEMENTS**

The work is performed according to the Russian Government Program of Competitive Growth of Kazan Federal University. Work was in part supported by the Russian Foundation for Basic Research (under the grants 14-07-00878, 15-37-21160).